\documentclass[10pt,twoside]{Lowx2011}
\usepackage{epsf,amsmath}
\usepackage{graphicx}

\begin{document}

\title{QCD ANALYSIS WITH DETERMINATION OF $\alpha_S(M_Z)$ BASED ON HERA INCLUSIVE AND JET DATA}

\author{\underline{A.M.COOPER-SARKAR} \\ \\
{\it for the H1 and ZEUS Collaborations} \\ \\
Dept. of Physics\\
Denys Wilkinson Laboratory\\
University of Oxford\\
Keble Rd, Oxford\\
OX1 3RH, UK\\
E-mail: a.cooper-sarkar@physics.ox.ac.uk}

\maketitle

\begin{abstract}
\noindent HERA jet data, in addition to combined HERA inclusive data, are input
 to the HERAPDF NLO QCD analysis in order to further constrain 
the gluon PDF and to allow an accurate evaluation of $\alpha_S(M_Z)$
\end{abstract}



\markboth{\large \sl \hspace*{0.25cm}\underline{A.M.Cooper-Sarkar} 
\hspace*{0.25cm} Low-$x$ Meeting 2011} {\large \sl \hspace*{0.25cm}QCD ANALYSIS WITH DETERMINATION OF $\alpha_S(M_Z)$ }

\section{Introduction}

The HERAPDF1.0 parton distribution functions (PDFs) were extracted using the 
combined inclusive cross section data from the H1 and ZEUS collaborations 
taken at the HERA collider during the HERA-I running period 
1992-1997~\cite{h1zeus:2009wt}. 
These data come from neutral and charged current interactions from both
$e^+ p$ and $e^- p$ scattering. The combination of the H1 and ZEUS data sets 
takes into account the full correlated systematic uncertainties of the 
individual experiments such that the total uncertainty of the combined measurement is typically smaller than 
$2\%$, for $3 < Q^2 < 500$~GeV$^2$, and reaches $1\%$, for $20 <  Q^2 < 
100$~GeV$^2$. These PDFs have been updated to HERAPDF1.5 
(see LHAPDFv6.8.6, http:projects.hepforge.org/lhapdf) by including a new 
preliminary combination of the HERA 
inclusive cross section data from the HERA-II running period 
2003-2007~\cite{herapdf15}. These data have greater accuracy at high $Q^2$ 
and high $x$ than the HERA-I data and thus they serve to decrease the PDF 
uncertainty in these kinematic regions.
In this presentation we discuss the extension of 
the HERAPDF1.5 analysis to include inclusive jet data. The new PDF set which 
results is called HERAPDF1.6~\cite{herapdf16}.
The jet data included in HERAPDF1.6 are: two sets of high-$Q^2$ 
inclusive jet production data from ZEUS~\cite{zeusjetdata}, low-$Q^2$ 
inclusive jet data from H1~\cite{h1lowq2jet} and normalised high-$Q^2$ 
inclusive jet data from H1~\cite{h1hiq2normjet}. 

The gluon PDF contributes only indirectly to the 
inclusive DIS cross sections. However, 
the QCD processes that give rise to scaling violations in the 
inclusive cross sections, namely the QCD-Compton (QCDC) and 
boson-gluon-fusion
(BGF) processes, are observed as events with distinct jets in the final 
state provided that the energy and momentum transfer are large enough.  The 
cross section for QCDC scattering depends on $\alpha_s(M_Z)$ and the quark 
PDFs. For HERA kinematics, 
this process dominates the jet cross section at large scales, 
where the quark densities are well known from the inclusive cross-section 
data, so that the value of $\alpha_s(M_Z)$ may be extracted without strong
correlation to the shape of the gluon PDF. The cross section for the
BGF process depends on $\alpha_s(M_Z)$ and the gluon PDF so that 
measurements 
of jet cross sections also provide a direct determination of the gluon 
density.

\section{Analysis}

Perturbative QCD predicts the $Q^2$ evolution of the parton distributions, but
not the $x$ dependence.
Parton distributions are extracted by 
performing a direct numerical integration of the DGLAP equations at NLO  
(NNLO fits are discussed in the presentation of Radescu).
A parametrised analytic shape for the parton distributions 
 is assumed to be valid  at some starting value of $Q^2 = Q^2_0$. 
For the HERAPDF the value $Q^2_0 = 1.9~$GeV$^2$ is chosen such 
that the starting scale is below the charm mass threshold, $Q_0^2 < m_c^2$. 
Then the DGLAP equations
are used to evolve the parton distributions up to a different $Q^2$ value, 
where they  are convoluted with NLO 
coefficient functions to make predictions 
for the structure functions.  The heavy quark 
coefficient functions are calculated in the general-mass
variable-flavour-number scheme of \cite{Thorne:1997ga}, with recent 
modifications~\cite{Thorne:2006qt}.
The heavy quark masses  for the central fit were chosen to be $m_c=1.4~$GeV 
and $m_b=4.75~$GeV and the strong coupling constant was fixed to 
$\alpha_s(M_Z) =  0.1176$. The predictions are then fitted to the 
combined HERA data sets for NC and CC $e^+p$ and $e^-p$ scattering. 
A minimum $Q^2$ cut of $Q^2_{min} = 3.5$~GeV$^2$ was imposed 
to remain in the kinematic region where
perturbative QCD should be applicable.

For the jet data the NLO cross-sections are made using NLOjet++~\cite{nlojet++}
but such calculations are too slow to be used in an iterative QCD fit hence 
the programme FASTNLO~\cite{fastnlo} is used to make these calculations by 
convoluting pre-calculated 
weights with the PDFs and $\alpha_S$.

 The fit parameters are those 
necessary to specify the input analytic shape.
Just as for HERAPDF1.0 and 1.5 the valence quark 
distributions $xu_v$, $xd_v$, and the $u$-type and $d$-type anti-quark 
distributions
$x\bar{U}$, $x\bar{D}$ ($x\bar{U} = x\bar{u}$, 
$x\bar{D} = x\bar{d} +x\bar{s}$) are parametrised at the input scale 
$Q^2_0=1.9$GeV$^2$ by the generic form 
\begin{equation}
 xf(x) = A x^{B} (1-x)^{C} (1 + D x + E x^2).
\label{eqn:pdf}
\end{equation}
The parametrisation for the gluon distribution $xg$ is extended, in 
comparison to the HERAPDF1.5 NLO parametrisation, to include a 
term $-A_{g'} x^{B_{g'}} (1-x)^{C_{g'}}$, such that the NLO gluon may become 
negative at low $x$ and low $Q^2$ (however it does not do so in the kinematic 
region of the HERA data). A further significant 
extension of the parametrisation is that 
the low-$x$ valence parameters, $B_{u_v}$ and $B_{d_v}$ 
are no longer set equal.
Otherwise the choice of central parametrisation is as for the published 
HERAPDF1.0(1.5) fits. Briefly,
the normalisation parameters, $A_g, A_{u_v}, A_{d_v}$, are constrained 
by the quark number sum-rules and momentum sum-rule. 
The $B$ parameters  $B_{\bar{U}}$ and $B_{\bar{D}}$ are set equal,
 $B_{\bar{U}}=B_{\bar{D}}$, such that 
there is a single $B$ parameter for the sea distributions. 
The strange quark distribution is expressed 
as $x$-independent fraction, $f_s$, of the $d$-type sea, 
$x\bar{s}= f_s x\bar{D}$ at $Q^2_0$.  The value $f_s=0.5$ would render the 
$s$ and $d$ quark densities the same, but the value $f_s=0.31$  
is chosen to be consistent with determinations 
of this fraction using neutrino-induced di-muon 
production. The further constraint 
$A_{\bar{U}}=A_{\bar{D}} (1-f_s)$, together with the requirement  
$B_{\bar{U}}=B_{\bar{D}}$,  ensures that 
$x\bar{u} \rightarrow x\bar{d}$ as $x \rightarrow 0$.
For the central fit only the $xu_v$ PDF has non-zero $E$ and $D$ parameters. 

The experimental uncertainties on the HERAPDF are determined using 
 the conventional 
$\chi^2$ tolerance, $\Delta\chi^2=1$, for $68\%$C.L. However model 
uncertainties and parametrisation uncertainties are also considered.
 The choice of the heavy 
quark masses is varied in the ranges,  
$1.35 < m_c < 1.65~$GeV and $4.3 < m_b <5.0~$GeV.
The choice of $Q^2_{min}$ is varied in the range, $2.5 < Q^2_{min} < 5.0$, and
 the choice of the strangeness fraction is varied in the range, 
$0.23 < f_s < 0.38$.
The difference between the central fit and the fits corresponding to model 
variations of $m_c$, $m_b$, $f_s$, $Q^2_{min}$ are 
added in quadrature, separately for positive and negative deviations, to
represent the model uncertainty of the HERAPDF1.6 set. 

Parametrisation variations for which
 the $E$ and the $D$ parameters for all the 
PDFs are freed one at a time are considered and variation of the 
starting scale $Q^2_0$ in the range, $1.5 < Q^2_0 < 2.5$GeV$^2$, is also 
considered as parametrisation variation. 
The difference between these parametrisation 
variations and the central fit is stored 
and an envelope representing the maximal deviation at each $x$ value
is constructed  
to represent the parametrisation uncertainty.

\section{Results}
\begin{figure}[htb]
\begin{tabular}{cc}
\includegraphics[width=0.48\textwidth]{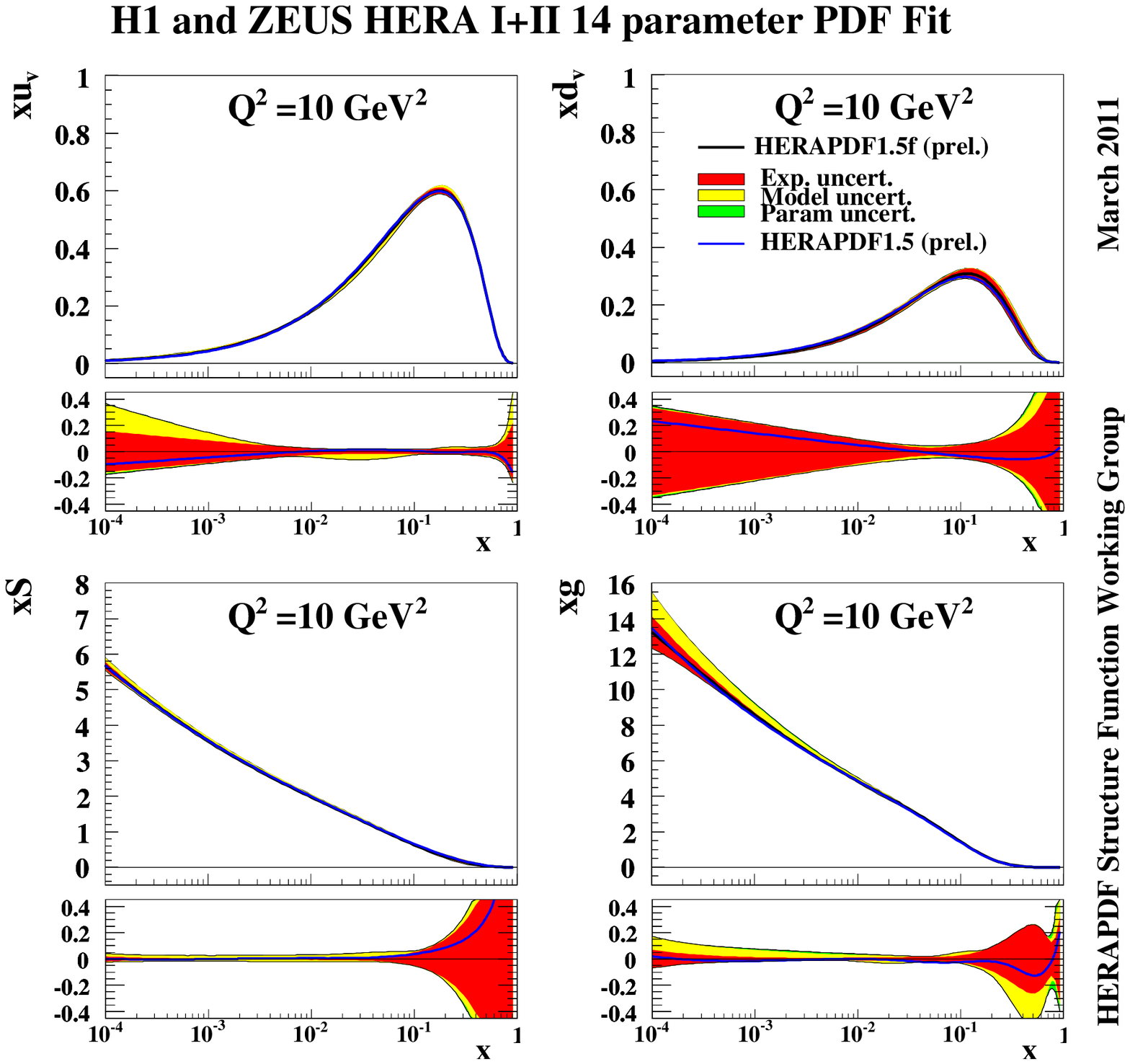} &
\includegraphics[width=0.48\textwidth]{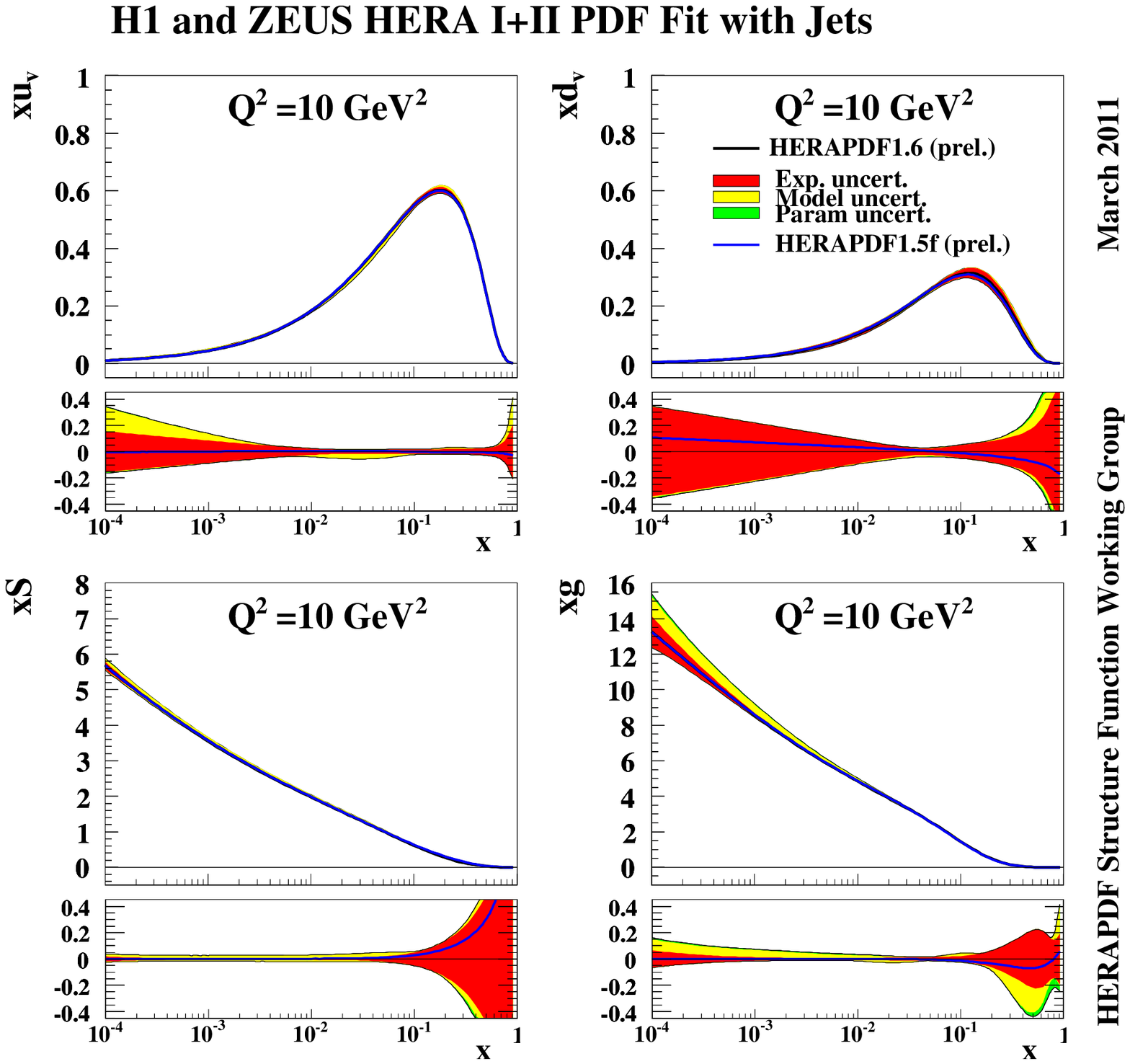}
\end{tabular}
\caption {The parton distribution functions 
$xu_v,xd_v,xS=2x(\bar{U}+\bar{D}),xg$, at $Q^2=10~$GeV$^2$, from
HERAPDF1.5f (left) and HERAPDF1.6 (right). 
Fractional uncertainty bands are shown 
below each PDF. The experimental, model and parametrisation 
uncertainties are shown separately. 
}
\label{fig:jetnojetfrac}
\end{figure}

\begin{figure}[htb]
\begin{tabular}{cc}
\includegraphics[width=0.45\textwidth]{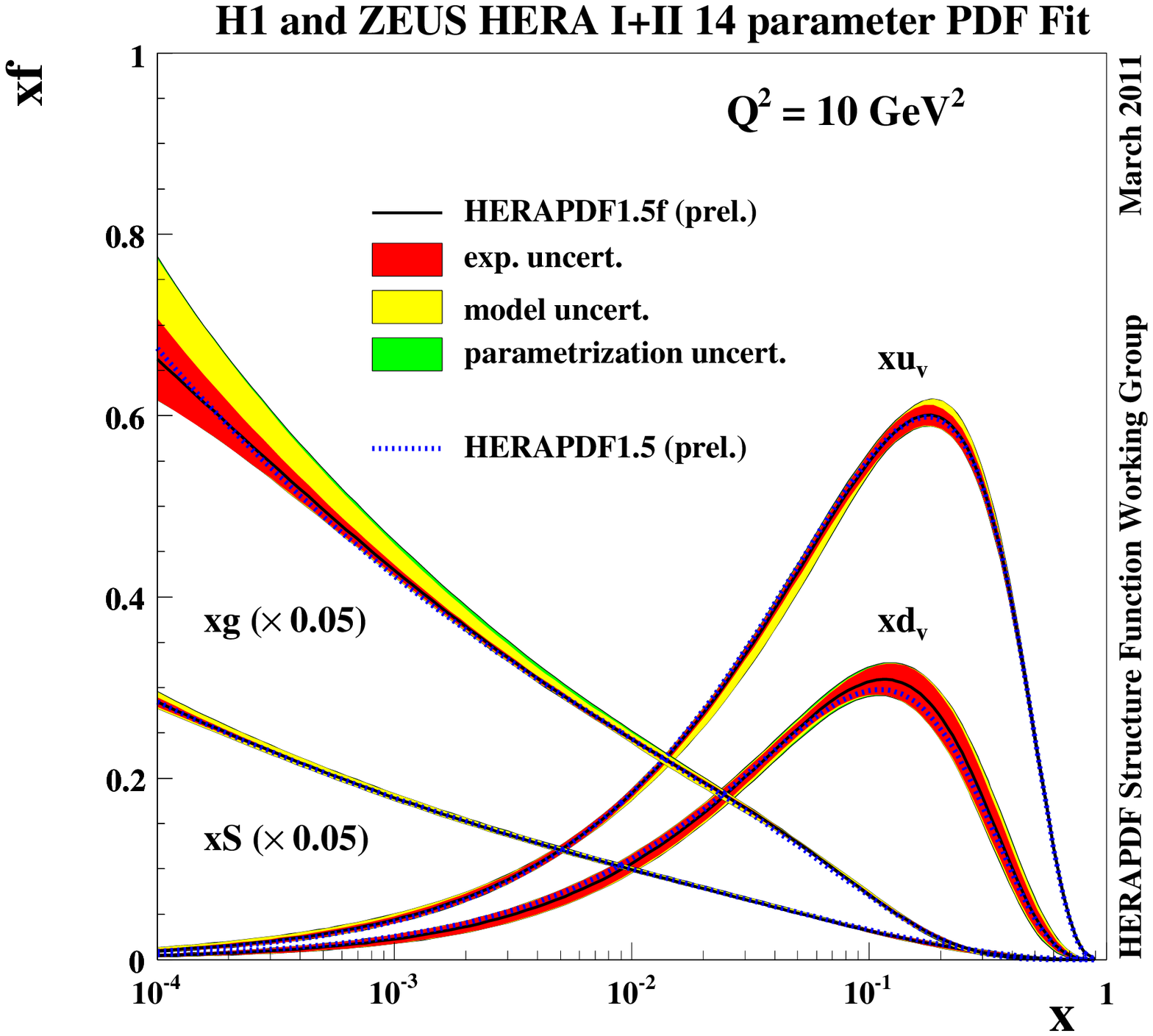} &
\includegraphics[width=0.45\textwidth]{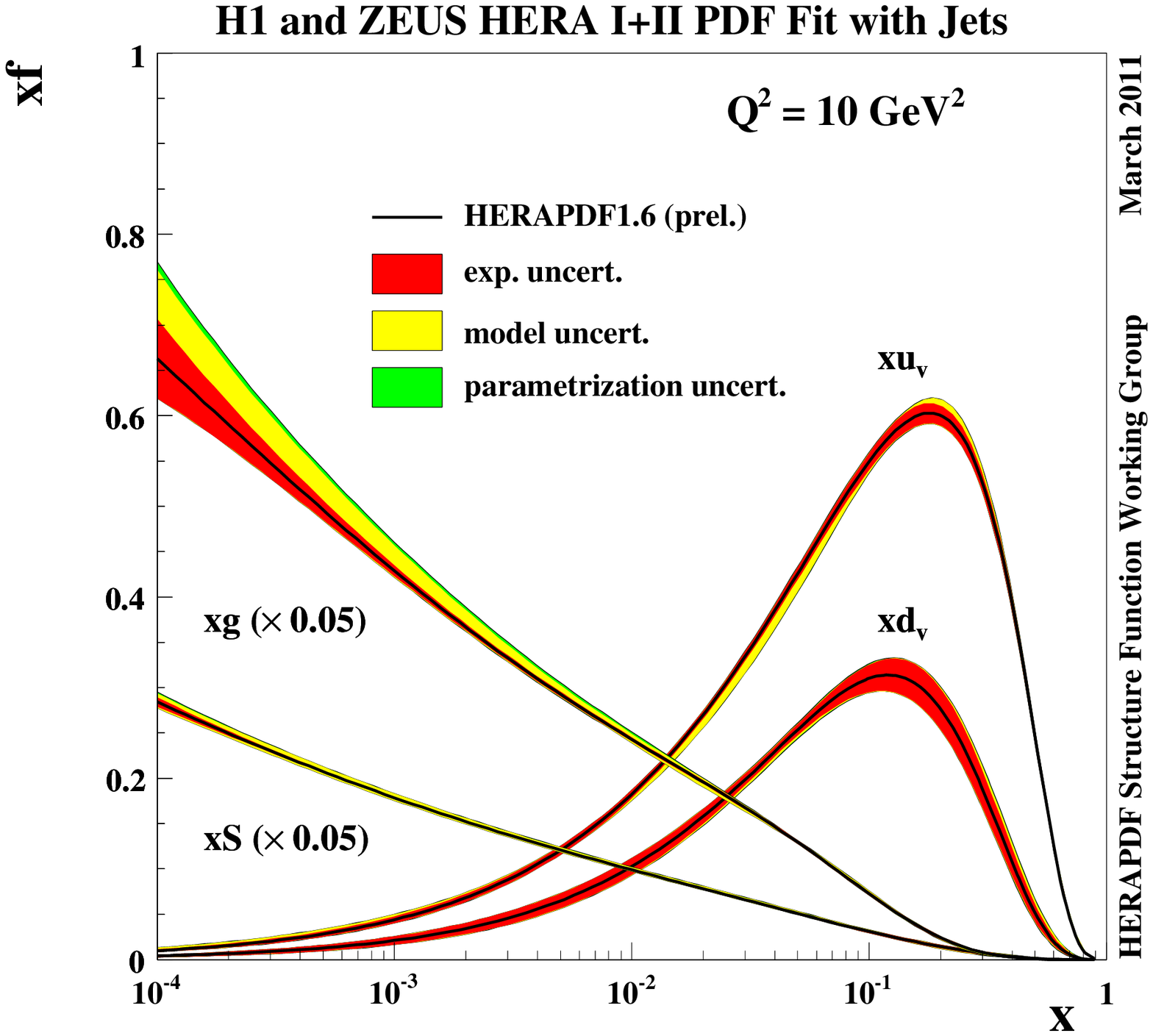}
\end{tabular}
\caption {The parton distribution functions 
$xu_v,xd_v,xS=2x(\bar{U}+\bar{D}),xg$, at $Q^2 = 10$~GeV$^2$ 
from HERAPDF1.5f and HERAPDf1.6.
The experimental, model and parametrisation 
uncertainties are shown separately. The gluon and sea 
distributions are scaled down by a factor $20$.
}
\label{fig:jetnojetsumm}
\end{figure}

Fig.~\ref{fig:jetnojetfrac} compares the HERAPDF1.6 NLO fit including the jet 
data (right) to a the HERAPDF1.5f NLO fit which does not include these data 
(left). Note that 
the HERAPDF1.5f fit differs from the publically available HERAPDF1.5 NLO 
fits in that it has the same 
flexible parametrisation for the central fit as HERAPDF1.6: 
there are extra parameters for 
the gluon PDF and the low-x behaviour of $u$-valence 
and $d$-valence are not required to be the same. 
The blue-line on the left-hand figure shows the 
HERAPDF1.5 central fit. Comparing this to the HERAPDF1.5f PDF shows that the 
extra flexibility has not changed the PDFs significantly.  
The blue line on the right-hand figure shows the central HERAPDF1.5f fit. 
Comparing this to the HERAPDF1.6 PDF shows that the addition of jets has not 
moved the PDFs outside their error bands, however the high-x Sea has become a 
little softer. 
Comparing the fractional uncertainty bands of the left and the right-hand 
plots shows that the addition of jets produces a marginal decrease in 
the high-$x$ gluon uncertainty. Fig.~\ref{fig:jetnojetsumm} compares the HERAPDF1.5f and HERAPDf1.6 fits in the form of summary of plots of $xu_v$,$xd_v$,$xg$ and the total Sea $xS=2x(\bar{U}+\bar{D})$.

The HERAPDF1.6 fit 
including jet data has a total $\chi^2$ of 811.5 for 780 data points (766 
degrees of freedom). For the inclusive data the $\chi^2$ is 730.2 for 674 data 
points and for the jet data the $\chi^2$ is 81.3 for 106 data points. 
For the HERAPDf1.5f fit without the inclusion of jet data the 
$\chi^2$ is 729 for 674 data points (660 degrees of freedom). 
Thus there is no tension between the jet data and the inclusive data.

\begin{figure}[htb]
\begin{tabular}{cc}
\includegraphics[width=0.45\textwidth]{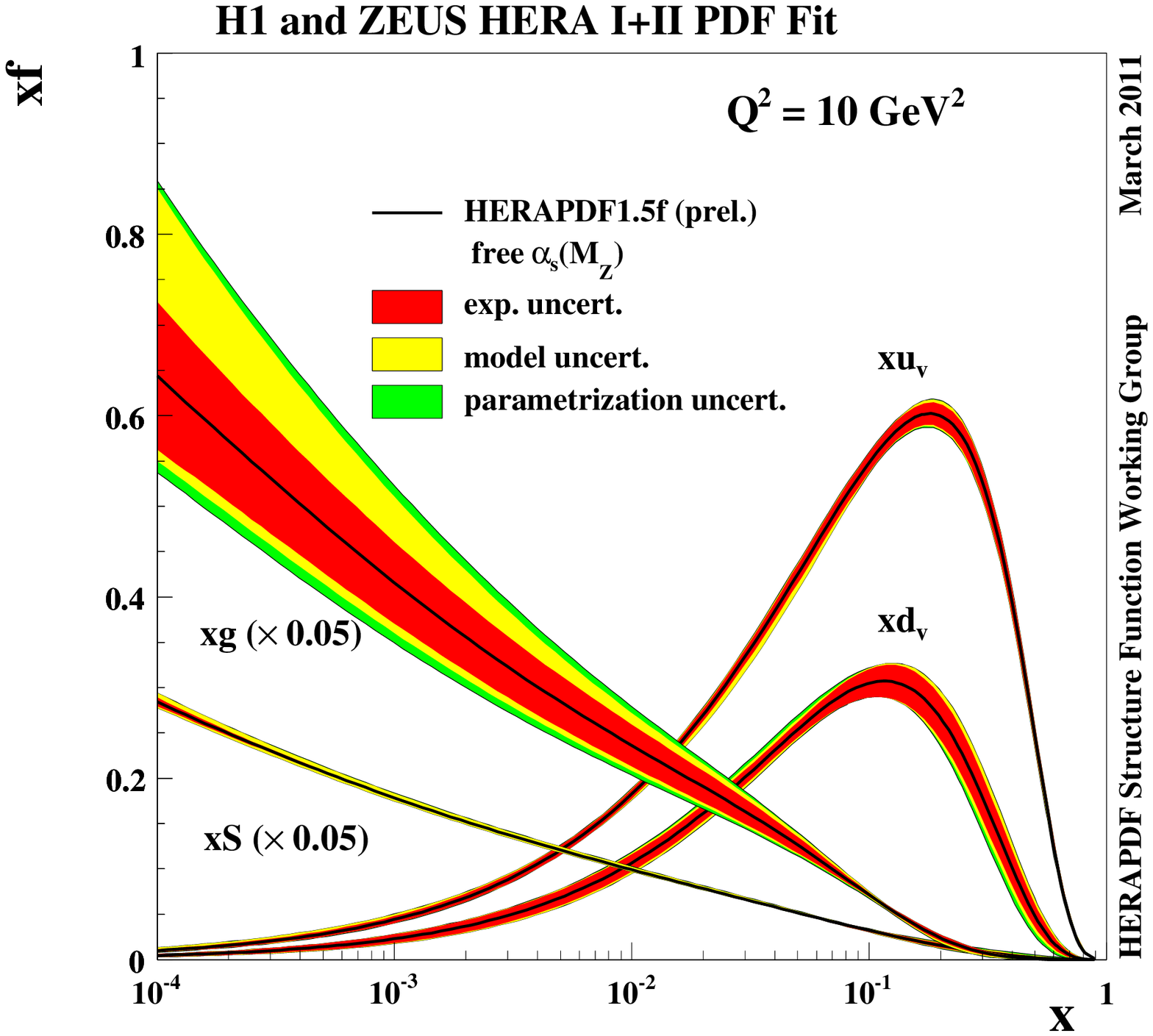} &
\includegraphics[width=0.45\textwidth]{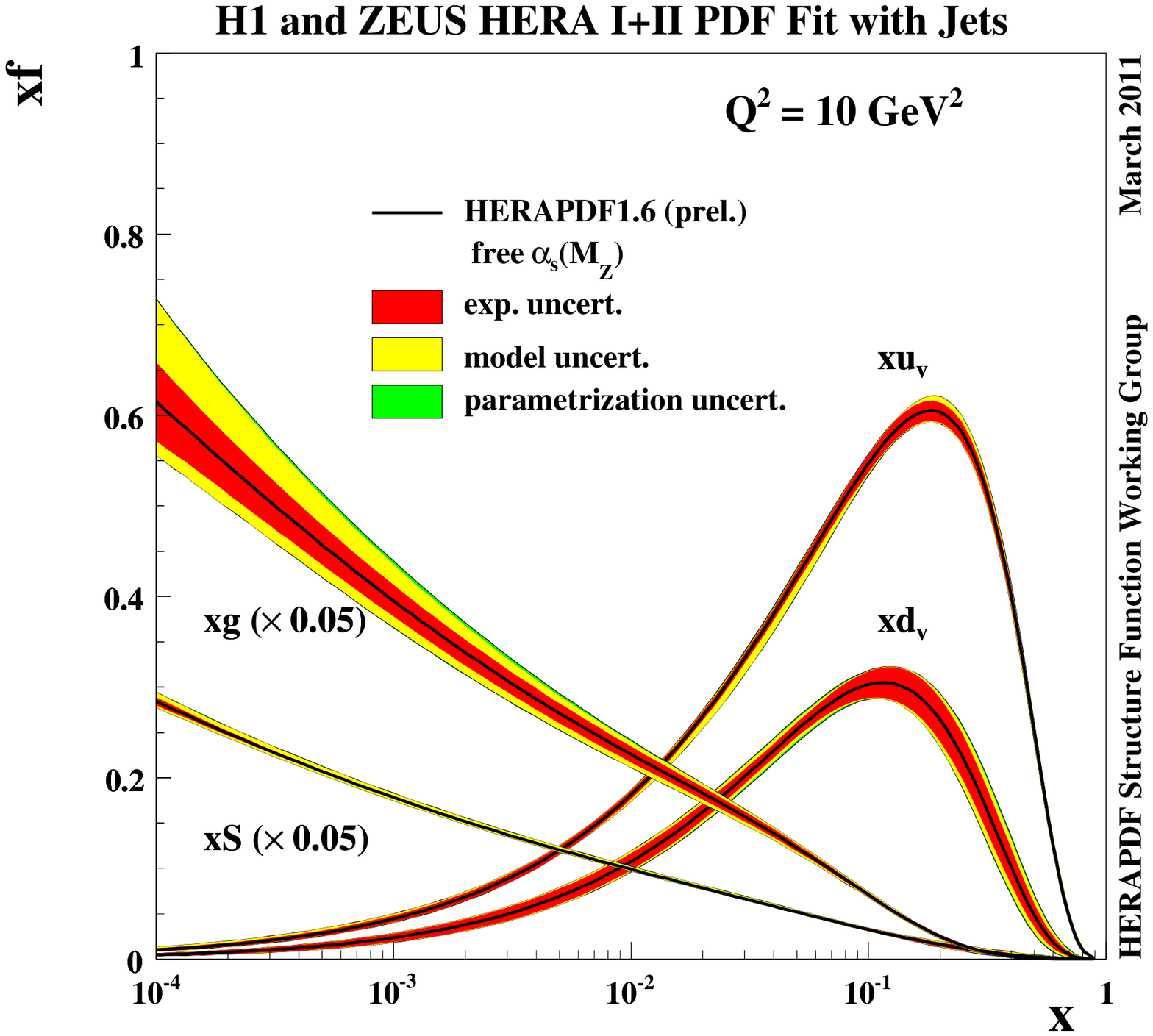}
\end{tabular}
\caption {The parton distribution functions
 $xu_v,xd_v,xS=2x(\bar{U}+\bar{D}),xg$, at $Q^2 = 10$~GeV$^2$, from 
HERAPDF1.5f and HERAPDf1.6, both with $\alpha_S(M_Z)$ 
treated as a free parameter of the fit.
The experimental, model and parametrisation 
uncertainties are shown separately. The gluon and sea 
distributions are scaled down by a factor $20$.
}
\label{fig:jetnojetalph}
\end{figure}

In most PDF analyses the strong coupling constant, $\alpha_S(M_Z)$, is fixed. 
However it can also be a parameter of the 
fit in addition to the PDF parameters. The $\chi^2$ for the HERAPDF1.6 fit 
with free $\alpha_S(M_Z)$ is 807.6 for 765 degrees of freedom. The sub-$\chi^2$
 for the inclusive data has barely changed but the sub-$\chi^2$ for the jet 
data decreases to 77.6 for 106 data points. 
Fig.~\ref{fig:jetnojetalph} shows the 
summary plots for HERAPDF1.5f and HERAPDF1.6, each with $\alpha_S(M_Z)$ left 
free in the fit. It can be seen that without jet data the uncertainty on 
the gluon PDF at low $x$ is large. This is because there is a strong 
correlation between the low-$x$ shape of
the gluon PDF and $\alpha_S(M_Z)$.  However once jet data are 
included the extra information on gluon induced processes reduces this 
correlation and the resulting 
uncertainty on the gluon PDF is not much larger than it 
is for fits with $\alpha_S(M_Z)$ fixed.

The value of $\alpha_s(M_Z)$ extracted from the HERAPDF1.6 fit is:

$
\alpha_S(M_Z) = 0.1202 \pm 0.0013(exp) \pm 0.0007(model/param) \pm 0.0012 (had) +0.0045/-0.0036(scale)
$

\begin{figure}[htb]
\begin{center} 
\includegraphics[width=0.4\textwidth]{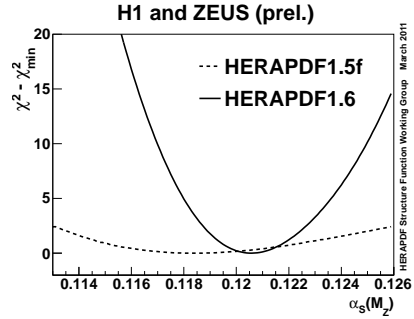} 
\caption {The difference between $\chi^2$ and its minimum value for the 
HERAPDF1.5f and HERAPDf1.6 fits as a function of $\alpha_s(M_Z)$
}
\end{center}
\label{fig:chiscan}
\end{figure}

We estimate the model and parametrisation uncertainties for $\alpha_S(M_Z)$ 
in the same way 
as for the PDFs and we also add the uncertainties in the hadronisation 
corrections applied to the jets. The scale uncertainties are estimated by 
varying the renormalisation and factorisation scales chosen in the jet 
publications by a factor of two up and down. The dominant contribution to the 
uncertainty comes 
from the jet renormalisation scale variation. 
Fig.~\ref{fig:chiscan}  shows a $\chi^2$ scan vs $\alpha_S(M_Z)$ for the fits 
with and without jets, illustrating how much better $\alpha_S(M_Z)$ is 
determined when jet data are included. The model and parametrisation errors 
are also much better controlled.

\begin{figure}[ht]
\begin{tabular}{cc}
\includegraphics[width=0.45\textwidth]{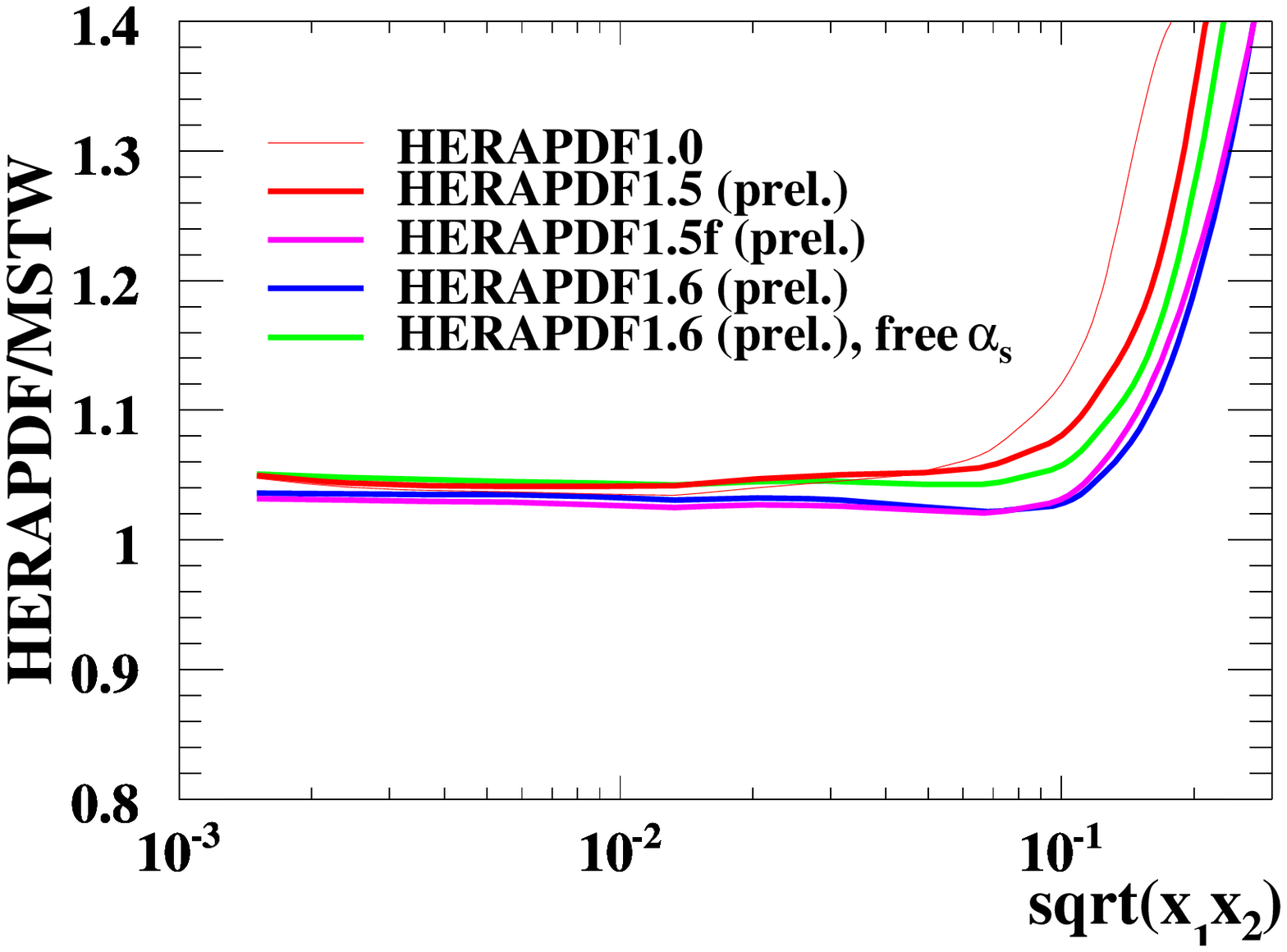} &
\includegraphics[width=0.45\textwidth]{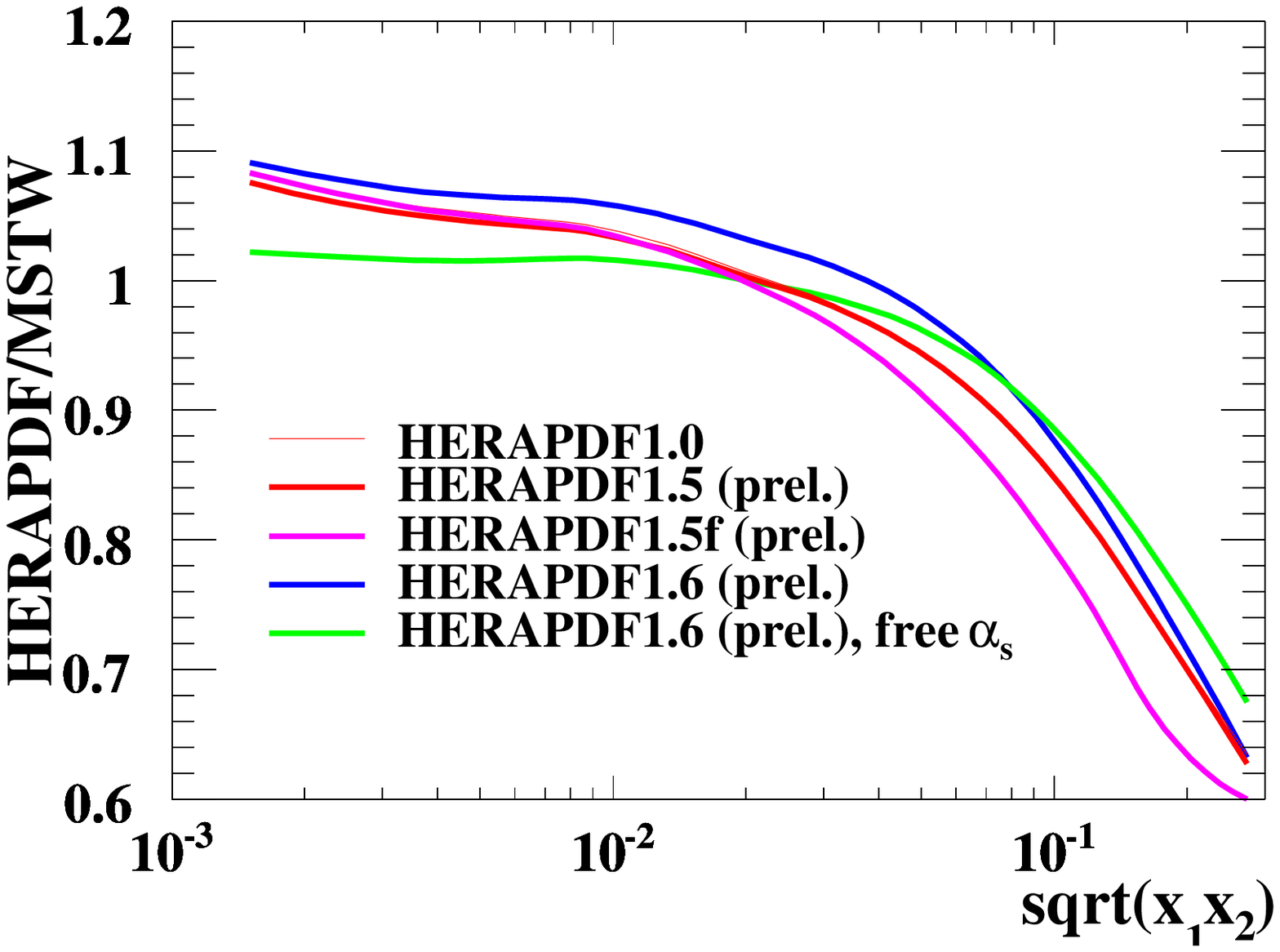}
\end{tabular}
\caption { The quark-anti-quark (left) and gluon-gluon (right) luminosities 
for the HERAPDFs in ratio to those for the MSTW2008 PDFs, as a function of the 
fraction of centre of mass energy taken by the sub-process, 
for the LHC at 7 TeV
}
\label{fig:lumi}
\end{figure}

Finally in Fig.~\ref{fig:lumi} 
we show plots of the gluon-gluon (right) and quark-antiquark (left) 
luminosities, 
as a function of the fraction of centre of mass energy taken by the subprocess,
for the LHC at 7 TeV, for all the NLO HERAPDF sets in ratio to the 
corresponding
luminosities extracted for the MSTW2008 PDFs. In general the HERAPDF1.5 and 
1.6 PDF sets are in better agreement with the MSTW2008 luminosities than 
the HERAPDF1.0 set. In 
particular the use of a larger value of $\alpha_S(M_Z)=0.1202$, for the 
HERAPDF1.6 free $\alpha_S(M_Z)$ PDF, brings the gluon-gluon luminosity into 
better agreement at low $x$ as well as at high $x$.

\section{Summary}
HERA jet data have been used in addition to HERA inclusive data in order to 
determine the parton distributions in the proton in an NLO QCD fit. 
There is no tension between the jet data and the inclusive data. The PDF set 
including the jet data, HERAPDF1.6, is similar to the publically available 
HERAPDF1.5 set both in PDF central values and in uncertainties. However, 
the advantage of using jet data is clearly seen when the strong coupling 
constant $\alpha_S(M_Z)$ is allowed to be a free parameter of the fits. 
The uncertainty on 
the low-$x$ gluon PDF due to the correlation with $\alpha_S(M_Z)$ is much 
reduced when jet data are used and an accurate value of $\alpha_S(M_Z)$ 
can be obtained:
$\alpha_S(M_Z) = 0.1202 \pm 0.0019$, excluding scale error.

\end{document}